# Crystal structure, site selectivity, and electronic structure of layered chalcogenide LaOBiPbS$_3$

Y. Mizuguchi[1], Y. Hijikata[1], T. Abe[2], C. Moriyoshi[2], Y. Kuroiwa[2], Y. Goto[1], A. Miura[3], S. Lee[4], S. Torii[4], T. Kamiyama[4], C. H. Lee[5], M. Ochi[6], K. Kuroki[6]

1. Graduate School of Science and Engineering, Tokyo Metropolitan University, 1-1, Minami-osawa, Hachioji 192-0397, Japan.
2. Department of Physical Science, Hiroshima University, 1-3-1 Kagamiyama, Higashihiroshima, Hiroshima 739-8526, Japan.
3. Faculty of Engineering, Hokkaido University, Kita-13, Nishi-8, Kita-ku, Sapporo, Hokkaido 060-8628, Japan.
4. Institute of Materials Structure Science, KEK, Tokai 319-1106, Japan.
5. National Institute of Advanced Industrial Science and Technology (AIST), Tsukuba, Ibaraki 305-8568, Japan.
6. Department of Physics, Osaka University, 1-1 Machikaneyama-cho, Toyonaka, Osaka 560-0043, Japan.

Abstract

We have investigated the crystal structure of LaOBiPbS$_3$ using neutron diffraction and synchrotron X-ray diffraction. From structural refinements, we found that the two metal sites, occupied by Bi and Pb, were differently surrounded by the sulfur atoms. Calculated bond valence sum suggested that one metal site was nearly trivalent and the other was nearly divalent. Neutron diffraction also revealed site selectivity of Bi and Pb in the LaOBiPbS$_3$ structure. These results suggested that the crystal structure of LaOBiPbS$_3$ can be regarded as alternate stacks of the rock-salt-type Pb-rich sulfide layers and the LaOBiS$_2$-type Bi-rich layers. From band calculations for an ideal (LaOBiS$_2$)(PbS) system, we found that the S bands of the PbS layer were hybridized with the Bi bands of the BiS plane at around the Fermi energy, which resulted in the electronic characteristics different from that of LaOBiS$_2$. Stacking the rock-salt type sulfide (chalcogenide) layers and the BiS$_2$-based layered structure could be a new strategy to exploration of new BiS$_2$-based layered compounds, exotic two-dimensional electronic states, or novel functionality.

Designing a new layered structure is one of the strategies for creating functional materials because low dimensional electronic states could be generated, and the electronic states could be easily tuned by altering stacking layer types. Among them, $BiS_2$-based compounds, whose crystal structure is composed of alternate stacks of an electrically conducting $BiS_2$ bilayer and various kinds of insulating layers (blocking layers), have recently drawn much attention as superconductors [1-11] and thermoelectric materials [12-16]. As were the Cu-oxide and Fe-based high-transition-temperature (high-$T_c$) superconductors [17,18], the superconducting properties and the thermoelectric performance of the $BiS_2$-based compounds, such as $RE(O,F)BiS_2$ (RE: rare earth), can be enhanced by replacing the elements at the blocking layer [19,20] and/or the conducting $BiS_2$ layers [21-24], which modifies the electronic states and local crystal structure [25-30]. As the strategies for designing new $BiS_2$-based compounds, explorations of new blocking layer structure or new types of conducting layers have been used, so far. Since the $BiS_2$-based compounds have a layered structure with van-der-Waals gap, making a stacking structure with ions or molecule layers, like intercalation, should be another strategy for creating new compounds with notable functions, as were FeSe-based, TNCl-based (T = Hf, Zr), and $Bi_2Se_3$ [31–35].

In this study, we have focused on a layered chalcogenide $LaOBiPbS_3$, which was firstly reported by Sun *et al.* in 2014 [15]. According to the previous report, the conducting layer is composed of rock-salt-type $M_4S_6$ layers with M = Bi and Pb; the M sites are randomly occupied by Bi and Pb. However, we noted that the bond lengths of M1-S and M2-S are apparently different in $LaOBiPbS_3$ (see Fig. 1 for the definitions of the M1 and M2 sites), which would indicate that the crystal structure of the $M_4S_6$ layer is not simple as described in Ref. 14. Therefore, we have analyzed the crystal structure of $LaOBiPbS_3$ using neutron diffraction and synchrotron X-ray diffraction. The structural refinements revealed site selectivity of Bi and Pb in the $LaOBiPbS_3$ structure. The Bi and Pb atoms preferably occupy the M1 and M2 sites, respectively. These results suggest that the crystal structure of $LaOBiPbS_3$ can be regarded as the alternate stacks of the rock-salt-type PbS-type layers and the $LaOBiS_2$-type layers, as shown in Fig. 1. Insertion of rock-salt-type chalcogenide layers into the van-der-Waals gap of $LaOBiS_2$ could be a new strategy to exploration of new $BiS_2$-based layered materials, exotic electronic states, or novel functionality.

The $LaOBiPbS_3$ polycrystalline samples were prepared by the solid-state reaction method. The $La_2O_3$ (99.9%), $La_2S_3$ (99.9%), $Bi_2S_3$, and PbS (99.9%) powders were mixed, pelletized, sealed into an evacuated quartz tube, and heated at 940 K for 25 hours. The obtained sample was ground, mixed to homogenize, pelletized, and heated at 1020 K for 25 hours. The phase purity of the prepared pellets was examined using laboratory X-ray diffraction (XRD) with Cu-$K_\alpha$ radiation. Neutron diffraction experiment was performed at room temperature by using a time-of flight type super-high-resolution powder diffractometer (SuperHRPD) [36-38] of the Japan Proton Accelerator Research Complex (J-PARC). The powder sample was installed into a Vanadium-Nickel Cell. We used the Z-Rietveld software [38] to perform Rietveld analysis. Synchrotron X-ray diffraction (SXRD) was performed at BL02B2 of SPring-8 with energy of 25 keV (project number: 2016B0074). The synchrotron XRD experiments were performed with a sample rotator system at room temperature, and the diffraction data were collected using a high-resolution one-dimensional semiconductor detector MYTHEN with a step of $2\theta = 0.006°$. The crystal structure parameters were refined using the Rietveld method with RIETAN-FP [39]. Schematic images of the crystal structure were drawn using VESTA [40].

Figure 2 displays the neutron diffraction pattern and the Rietveld refinement results for the LaOBiPbS$_3$ sample. The obtained pattern could be refined using the structural model with the tetragonal $P4/nmm$ space group (No. 129) with small impurities of La$_2$O$_2$S and PbS. The refined lattice constants were $a = 4.09676(1)$ Å and $c = 19.7996(6)$ Å. Assuming that concentrations of Bi and Pb in LaOBiPbS$_3$ are the same, the refined occupancies of Bi for M1 and M2 were converged with 0.89 and 0.11, respectively: Bi:Pb = 89(2):11(2) for M1 and Bi:Pb = 11(3):89(3) for M2 are obtained. The final reliability factor $R_{wp}$ was 5.16%. The refinement with the random model, in which M1 and M2 site was a fixed occupancy of Bi:Pb = 50:50, resulted in slightly higher reliability factor: $R_{wp}$ = 5.19%. Indeed, the M1 site is selectively occupied by Bi, while the M2 site is selectively occupied by Pb in LaOBiPbS$_3$.

Figure 3 shows the SXRD pattern and the Rietveld refinement result for the LaOBiPbS$_3$ sample. The obtained pattern was refined using a structural model with the tetragonal $P4/nmm$ space group, as well. Impurity phases of La$_2$O$_2$S (11%) and PbS (7%) were detected. The refined lattice constants were $a = 4.09717(4)$ Å and $c = 19.7933(2)$ Å. The reliability factor $R_{wp}$ was 7.5%. Note that the occupancies for M1 and M2 were fixed as the values obtained from neutron diffraction in the final Rietveld refinement of the SXRD pattern presented in Fig. 3. The refined lattice constants and atomic positions were close to those determined by neutron diffraction. The crystal structure parameters obtained from neutron diffraction and SXRD were listed in Tab. 1. In Tab. 2, the typical bond lengths and the bond angles around the M1 and M2 sites are listed. From the refined bond lengths, we calculated bond valence sum for the M1 and M2 sites as 3.17 and 2.19, respectively. These values of bond valence sum also suggest that the M1 and M2 sites are selectively occupied by $Bi^{3+}$ and $Pb^{2+}$, respectively, which is consistent with that obtained from neutron diffraction.

To visualize the refined structure, the typical bond lengths and bond angles are displayed with a crystal structure image of $LaOBiPbS_3$ in Fig. 4. The M1 site is surrounded by S1 and S2, forming Bi-rich $M1S_5$ pyramids, in which M1 and S2 locate the bottom center and the apical position, respectively. The Bi-rich $M1S_5$ pyramid structure is quite similar to that in $LaOBiS_2$ [30]: shorter M1-S2 bonds and longer M1-S1 bonds form the pyramid. The M2-S3 bonding forms rock-salt-like layers. As shown in Fig. 4, the in-plane M2-S3 bond length and the M2-S3 bond length along the $c$-axis direction are 2.8987 and 2.9443 Å, respectively. From the almost isotropic bonding in the M2-S3 layers, the structure can be regarded as a distorted rock-salt structure. In fact, the bond length of Pb-S in rock-salt-type PbS is 2.989 Å [41], which is close to the M2-S3 lengths in $LaOBiPbS_3$. From these structural characteristics, the structure of $LaOBiPbS_3$ can be described as Bi-rich $LaOM1S_2$ sandwiched by Pb-rich M2S layer. The in-plane (M1-S1) bond length in $LaOBiPbS_3$ is slightly longer than the in-plane Bi-S1 distance in $LaOBiS_2$, which can be explained by the misfit between Bi-rich M1-S1 and Pb-rich M2-S2.

On the basis of the refined crystal structure of $LaOBiPbS_3$, we proposed that the crystal structure of $LaOBiPbS_3$ could be regarded as the alternate stacks of PbS-type (rock-salt-type) and $LaOBiS_2$-type layers, as shown in Fig. 1. Insertion of rock-salt-type sulfide layers into the $BiS_2$-type layered structures, should be a concept useful for designing new $BiS_2$-based superconductors, thermoelectric materials, and functional materials. Since there are many kinds of $BiS_2$-based compounds with various types of blocking layers, the material design concept presented here will expand the playground of chemistry and physics of $BiS_2$-based functional materials.

To understand the effects of the insertion of PbS-type layers into the $LaOBiS_2$-type layers on the electronic states, we have calculated the band structure of the ideal $LaOBiPbS_3$ structure with 100% occupancy (complete site selectivity) of Pb at the M2 site, such as $(LaOBiS_2)(PbS)$. Figure 5 presents the first-principles band structure and the (partial) density of states calculated with the WIEN2k code [42] using the Perdew-Burke-Ernzerhof parametrization of the generalized gradient approximation (GGA-PBE) [43]. We found that the S bands of the PbS layers are hybridized with the Bi bands of BiS plane around the Fermi energy. The band edge of the Pb bands and the S bands of the BiS plane near the X point are located at around 1.2 eV above and 1 eV below the Fermi energy, respectively. The calculated electronic structure with density of states at Fermi energy looks quite different from LaOBiS$_2$, although it is not decisive whether LaOBiPbS$_3$ (with 100% occupancy of Pb at the M2 site) is metallic since the GGA-PBE approximation is known to underestimate the band gap. We expect, however, tunable band structure by the insertion of various kinds of chalcogenide layers, like PbS into the van-der-Waals gap of the BiS$_2$-based compounds. Thus, stacking the rock-salt-type sulfide (chalcogenide) layers and the BiS$_2$-based layered structure should be a new strategy of band engineering for designing new BiS$_2$-based layered superconductors, thermoelectric materials, or novel functional materials.

In conclusion, the crystal structure of LaOBiPbS$_3$ has been examined using neutron powder diffraction and powder synchrotron X-ray diffraction. The structural refinements revealed that two metal sites, M1 and M2 sites, were differently surrounded by sulfur atoms. Furthermore, we found site selectivity of Bi and Pb in the LaOBiPbS$_3$ structure. The structural characteristics around the M1 and the M2 site are close to the Bi site of LaOBiS$_2$ and the M2 site are close to the Pb site of rock-salt-type PbS, respectively. These results suggested that the crystal structure of LaOBiPbS$_3$ can be regarded as alternate stacks of the Pb-rich PbS-type layers and the Bi-rich LaOBiS$_2$-type layers. From band calculations for an ideal (LaOBiS$_2$)(PbS) system, we found that the S bands of the PbS layer are hybridized with the Bi bands of the BiS plane at around the Fermi energy, which results in the electronic characteristics different from that of LaOBiS$_2$. Thus, stacking the rock-salt type sulfide (chalcogenide) layers and the BiS$_2$-based layered structure can largely modify the electronic structure and should be a new strategy of band engineering for designing new BiS$_2$-based layered superconductors, thermoelectric materials, or novel functional materials.

Acknowledgements
    The authors thank O. Miura for his experimental support. This work was partly supported by Grant-in-Aid for Scientific Research (Nos. 15H05886, 16H04493, and 17H05481) and JST-CREST (No. JPMJCR16Q6), Japan. The neutron diffraction experiments using SuperHRPD were carried out under the S-type project of IMSS, KEK with Proposal No.2014S05. The SXRD experiments were performed with Proposal No. 2016B0074 at SPring-8.

Table 1. Crystal structure parameters for LaOBiPbS$_3$ obtained from neutron diffraction and synchrotron X-ray diffraction (SXRD).

| Site | Occupancy (neutron) | $x$ | $y$ | $z$ (neutron) | $z$ (SXRD) | $U$ (neutron) (Å$^2$) | $U$ (SXRD) (Å$^2$) |
|---|---|---|---|---|---|---|---|
| La | 1 | 0 | 0.5 | 0.06184(3) | 0.06184(9) | 0.0014(2) | 0.0029(5) |
| O | 1 | 0 | 0 | 0 | 0 | 0.0083(2) | 0.013(fixed) |
| M1 | Bi$_{0.89(2)}$Pb$_{0.11(2)}$ | 0.5 | 0 | 0.57638(3) | 0.57681(7) | 0.0104(2) | 0.0115(4) |
| M2 | Bi$_{0.11(3)}$Pb$_{0.89(3)}$ | 0.5 | 0 | 0.25805(4) | 0.25784(8) | 0.0077(2) | 0.0126(4) |
| S1 | 1 | 0.5 | 0 | 0.72980(9) | 0.7293(4) | 0.0093(5) | 0.010(2) |
| S2 | 1 | 0.5 | 0 | 0.4290(1) | 0.4281(4) | 0.0059(5) | 0.012(2) |
| S3 | 1 | 0.5 | 0 | 0.1342(1) | 0.1349(4) | 0.0067(4) | 0.017(2) |

Table 2. Typical bond lengths and angles of LaOBiPbS$_3$ determined from SXRD analysis.

| | |
|---|---|
| M1-S1 | 2.9083(8) Å |
| M1-S2 | 2.433(9) Å |
| M1-S3 | 3.370(9) Å |
| M2-S1 | 3.018(9) Å |
| M2-S3 (in-plane) | 2.8988(3) Å |
| M2-S3 ($c$-axis) | 2.943(9) Å |
| S1-M1-S1 angle | 170.0(3)° |
| S3-M2-S3 angle | 176.2(3)° |

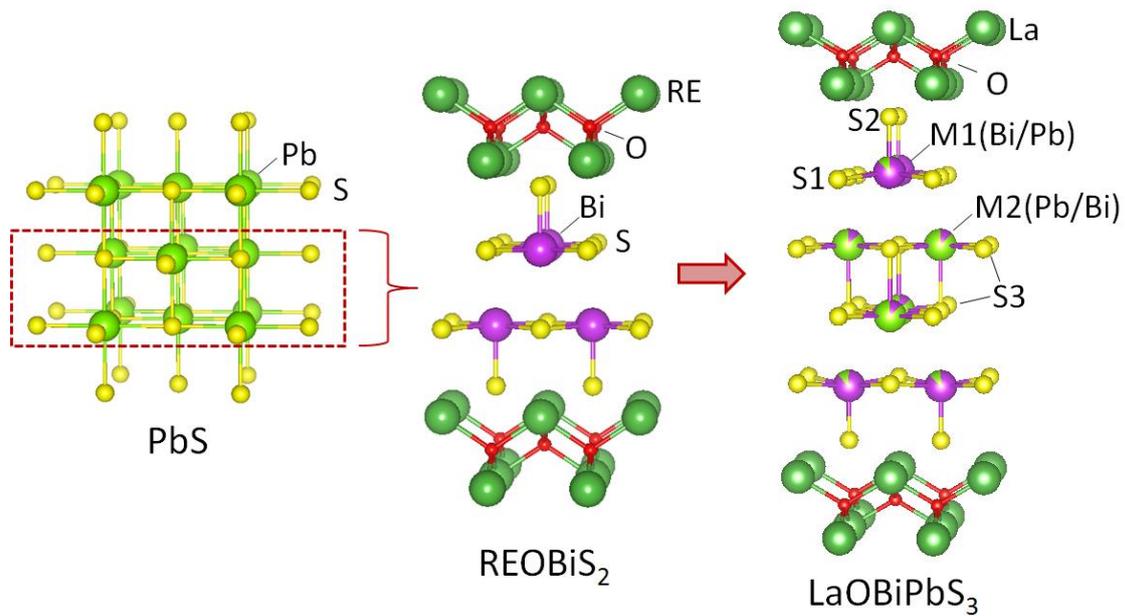

Fig. 1. Schematic images of the relationship of the crystal structure between PbS (rock-salt-type cubic), REOBiS$_2$ (tetragonal), and LaOBiPbS$_3$.

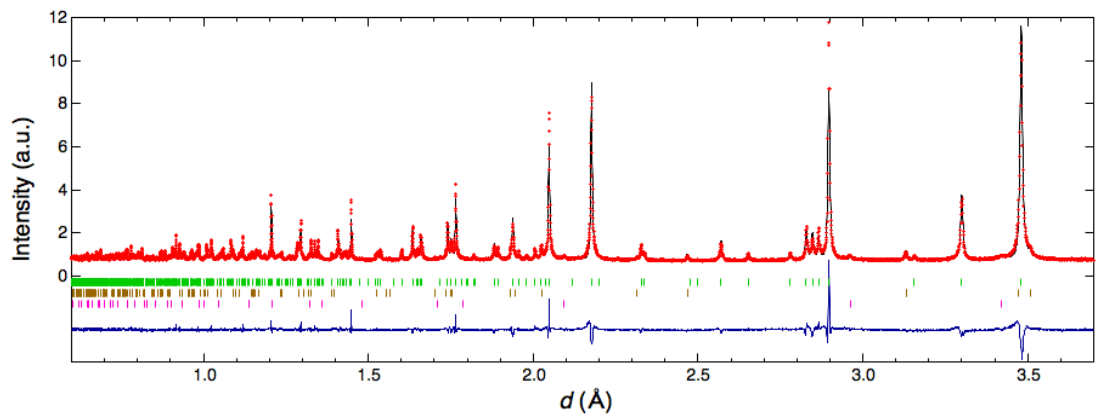

Fig. 2. Neutron diffraction pattern and refinement results for the LaOBiPbS$_3$ sample.

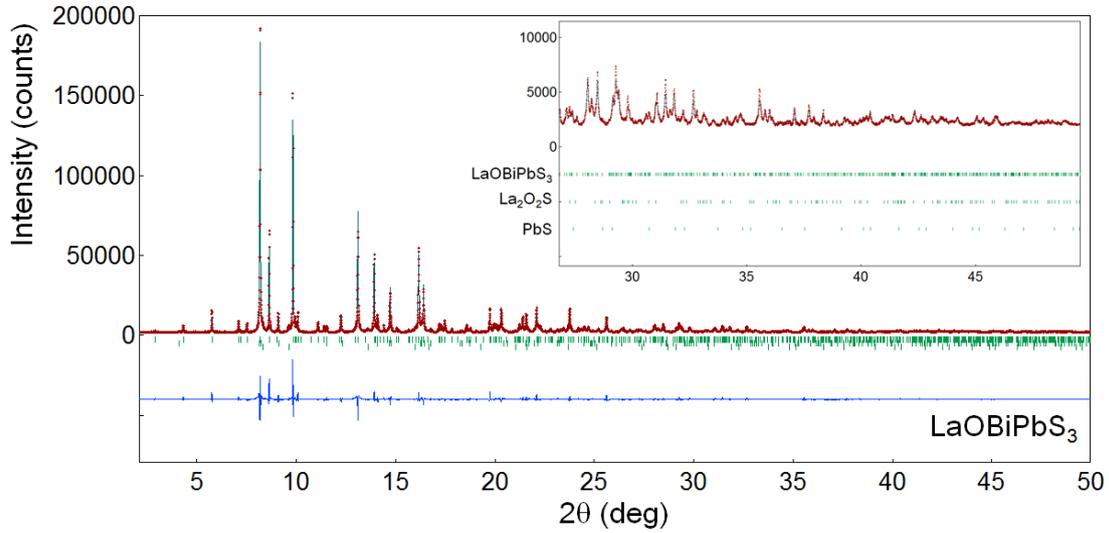

Fig. 3. SXRD pattern and the result of Rietveld analysis for the LaOBiPbS$_3$ sample. The Rietveld analysis was performed by multi-phase refinement with the La$_2$O$_2$S and PbS impurity phases. The inset shows an enlarged profile at higher angles.

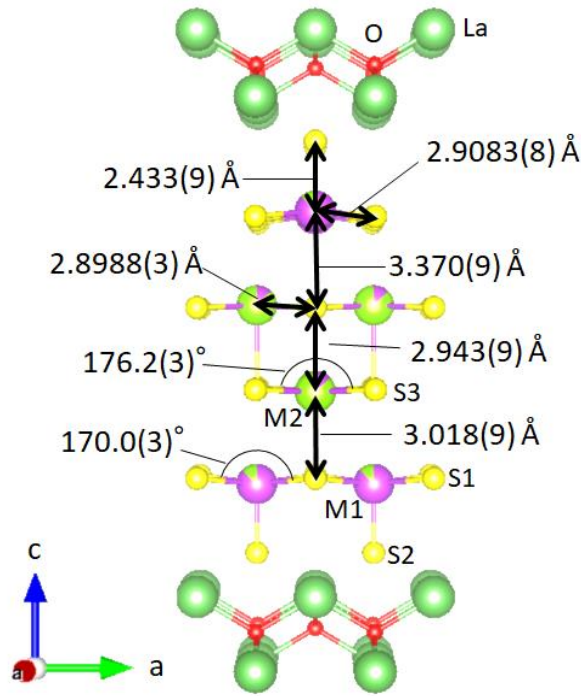

Fig. 4. Schematic image of LaOBiPbS$_3$ with the typical bond lengths and angles determined from the structural refinements. M1 is Bi-rich site while M2 is Pb-rich one.

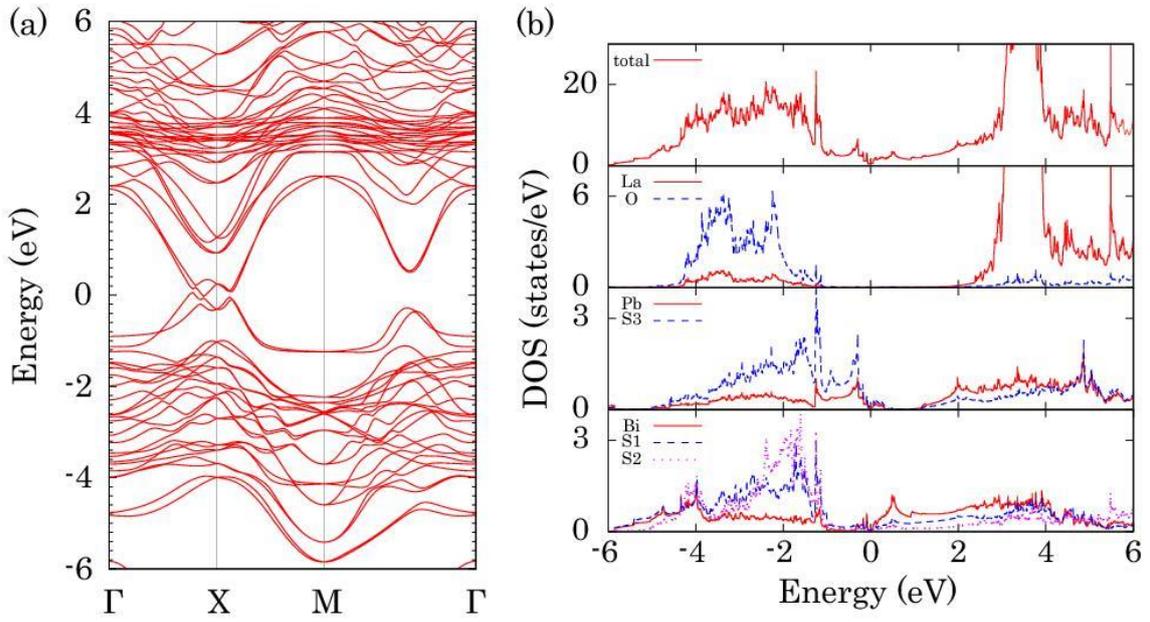

Fig. 5. (a) Band structure and (b) partial density of states for LaOBiPbS$_3$ with the ideal structure (M1 = Bi, M2 = Pb).